# Rapidly-converging multigrid reconstruction of cone-beam tomographic data

Glenn R. Myers[a], Andrew M. Kingston[a], Shane J. Latham[a], Benoit Recur[b], Thomas Li[a], Michael L. Turner[a], Levi Beeching[a], and Adrian P. Sheppard[a]

[a]The Australian National University, ACT, Australia [b]Institut National de la Santé et de la Recherche Médicale, France

**ABSTRACT**

In the context of large-angle cone-beam tomography (CBCT), we present a practical iterative reconstruction (IR) scheme designed for rapid convergence as required for large datasets. The robustness of the reconstruction is provided by the "space-filling" source trajectory along which the experimental data is collected. The speed of convergence is achieved by leveraging the highly isotropic nature of this trajectory to design an approximate deconvolution filter that serves as a pre-conditioner in a multi-grid scheme. We demonstrate this IR scheme for CBCT and compare convergence to that of more traditional techniques.

## 1. INTRODUCTION

In X-ray computed tomography (CT), radiographs are collected as the X-ray source and detector move relative to the sample. From this sequence of radiographs, CT reconstruction methods use a model of the tomographic imaging system to reconstruct a 3d image of the sample.[1] The resolution of the reconstructed 3d image is limited by: the information content of the radiographs; the quality (i.e. signal-to-noise ratio) of the recorded radiographs; and the accuracy of the assumed model of the tomographic imaging system.

The information content of the radiographs is sufficient to solve the reconstruction problem if appropriate scanning parameters are chosen.[1] The X-ray energy and flux must be chosen such that attenuation contrast is visible at the detector, and the radiographs must be taken at a sequence of points on a scanning trajectory that satisfies appropriate data completeness conditions.[1] Helical and space-filling trajectories are both examples of "complete" scanning trajectories.[2-4]

Noise in the radiographs leads to noise in the reconstructed 3d image.[1] In contrast, inaccuracies in the assumed model of the tomographic imaging system can lead to more structured artefacts.[5-10] For example, if the component of the imaging system deviate from their assumed locations (e.g due to thermal drift), this manifests as blur or doubled edges in the sample.[5-10] Incorrect modeling of refraction or spectral effects can lead to false edges and streaking.[11] These structured artifacts are particularly problematic for automated interpretation of tomographic images, which is becoming increasingly necessary as these images becomes larger (see, e.g. the 186 GigaVoxel dataset presented in Myers et. al. 2016[12]).

In order to reduce structured artefacts and improve the predictive power of CT imaging methods, we thus wish to improve our assumed models of the tomographic imaging system. This sort of parameter estimation can either be done as a pre-processing step before non-iterative tomographic reconstruction, or jointly with iterative reconstruction of the tomographic image. Iterative model correction as a pre-processing step leads to substantial improvements in image quality, but is becoming increasingly computationally costly. Many correction techniques require a partial, or low-resolution reconstruction of the 3d image.[5-10] This motivates us to explore iterative CT reconstruction methods, where the model parameters for the tomographic imaging system may be estimated jointly with the 3d image itself.[6,7,9,10,13]

___

A variety of iterative CT reconstruction methods are present in the literature. A "gold standard" for simplicity and robustness is Landweber iteration [known in the context of CT as the simultaneous iterative reconstruction technique (SIRT)], whilst more complex Bayesian reconstruction algorithms [e.g. expectation maximisation for transmission tomography (EMTR)] can display faster convergence and use more physically accurate noise models.[14] Although these algorithms allow model parameter estimation to proceed jointly with the reconstruction[9,13] they are not used routinely due to their prohibitive computational cost. Iterative CT reconstruction methods take tens to hundreds of iterations to converge, and require at least one projection and back-projection operation per iteration.[14] These projection and back-projection operations have computational complexity scaling as $O(N^3 \ln N)$ (or $O(N^4)$ in most implementations), where $N$ is the number of voxels along one side of the 3d reconstructed image, and so are extremely time-consuming.[15,16] In contrast, non-iterative methods require only a single back-projection operation.

In order to make iterative CT reconstruction competitive for routine imaging it is thus essential to explore methods to speed convergence, and reduce the number of iterations required to produce an acceptable 3d image. In this paper we present a method for preconditioned multigrid Landweber iteration, and show that it produces an acceptable 3d reconstructed image in only two iterations. The computational load of this method is thus close to that of non-iterative reconstruction methods. We compare the reconstruction quality and convergence speed to SIRT, and a variant of EMTR where convergence has been accelerated using ordered subsets. The extremely fast convergence of this method comes largely from the preconditioning, which is possible when using a scanning trajectory with an approximately spatially invariant point-spread function (PSF), such as a spacefilling trajectory.[4]

## 2. BENCHMARK ITERATIVE RECONSTRUCTION TECHNIQUES

To illustrate the performance (both in terms of quality and convergence) of the preconditioned multigrid Landweber solver, we will use a micro-CT dataset collected by Grzegorz Pyka at FEI. A micro-focus X-ray source was used with an accelerating voltage of 70keV and source current of 90$\mu$A; low energies were filtered from the beam using 0.1mm of steel placed over the source aperture. A geopolymer sample and Varian flat-panel detector were placed 7.52mm and 330mm from the source point respectively. Using 0.7s exposures, 11099 radiographs ($3040^2$ pixel) were collected along a space-filling trajectory with a stride of 15.3 degrees.[4] As such a large dataset would take an unacceptably long time to reconstruct using conventional iterative techniques, a 3600 radiograph subset of the data was extracted and downsampled by a factor of 4. Reconstruction was performed using a GPU-accellerated, MPI-parallel framework,[12] on a desktop box with 512GB of RAM, 2 Intel E5-2690v3 12-core CPUs, and 3 nVidia Titan X GPUs (each with 12GB onboard RAM). The resulting volume was 640×640×2304 voxels.

Making the projection approximation and ignoring refraction, we express this CT imaging process mathematically as

$$\langle I(\mathbf{r},\Theta) \rangle = I_0(\mathbf{r},\Theta) \exp\left[-(\mathcal{P}\mu)(\mathbf{r},\Theta)\right] \quad (1)$$

In this notation the vector **r** is a 2d position on the detector, Θ specifies a source and detector position along the scanning trajectory, $\langle I \rangle$ is the expected value of the X-ray intensity recorded by the detector, $I_0$ is the X-ray intensity incident on the sample, P is the X-ray projection operator, and $\mu(\mathbf{x})$ is the linear attenuation coefficient of the sample at 3d Cartesian coordinates **x**.

### 2.1 Simultaneous Iterative Reconstruction Techniques (SIRT)

SIRT reconstruction is a straightforward Landweber iteration according to[14]

$$\mu^{(n+1)}(\mathbf{x}) = \mu^{(n)}(\mathbf{x}) + \alpha \mathcal{B}\left\{\ln\left[\frac{I_0(\mathbf{r},\Theta)}{I(\mathbf{r},\Theta)}\right] - [\mathcal{P}\mu^{(n)}](\mathbf{r},\Theta)\right\}, \quad (2)$$

where B is the back-projection operator, (adjoint to the projection operator P), $\alpha$ is a tuning parameter typically equal to the detector width in pixels, and $\mu^{(n)}$ is the estimate of $\mu$ at the $n$th iteration. Each iteration requires a single projection and back-projection operation. A reconstruction of the geopolymer sample is shown in figure 1. The blur in the images demonstrates that the reconstruction algorithm has failed to converge after 10 iterations.

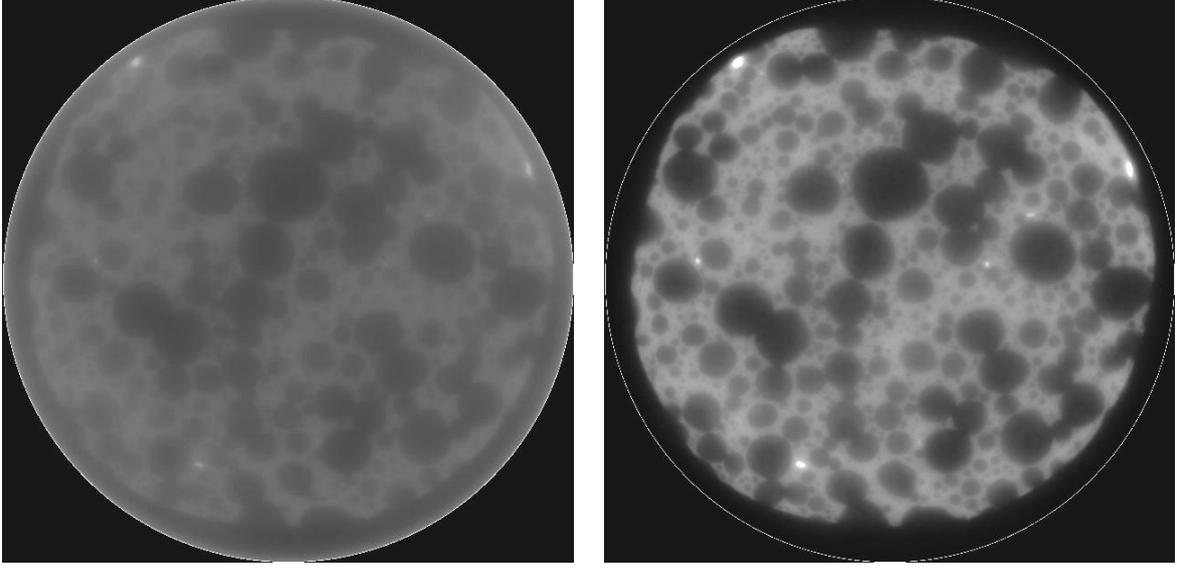

Figure 1. A $640^2$ voxel slice through the 3d SIRT reconstruction of a geopolymer sample, taken orthogonal to the rotation axis, after 2 (left) and 10 (right) iterations. The blurry areas are where the reconstruction has yet to converge.

## 2.2 Ordered-Subset Expectation Maximisation for TRansmission tomography (EMTR)

As a second point of comparison we consider EMTR with ordered subsets (OS). This algorithm is a Bayesian maximum likelihood estimation algorithm that may be obtained by assuming the recorded contrast ln[$I(\mathbf{r},\Theta)$] follows a Poisson probability distribution.[17] To accelerate convergence the radiographs are divided into $M$ disjoint subsets denoted here as $I(\mathbf{r},\Theta^{(m \in M)})$. We then define a single full-resolution iteration as involving a single pass over every subset according to:

$$\mu^{(n+1/M)}(\mathbf{x}) = \mu^{(n)}(\mathbf{x}) + \frac{\mathcal{B}\left\{\exp\left[-[\mathcal{P}\mu^{(n)}](\mathbf{r},\Theta^m)\right] - \frac{I(\mathbf{r},\Theta^m)}{I_0(\mathbf{r},\Theta^m)}\right\}}{\mathcal{B}\left\{\exp\left[-[\mathcal{P}\mu^{(n)}](\mathbf{r},\Theta^m)\right] \times (\mathcal{PI})(\mathbf{r},\Theta^m)\right\}}, \quad (3)$$

where PI denotes the length of the projection of the support of $\mu(\mathbf{x})$. OS-EMTR reconstructions of the geopolymer sample are shown in figure 2, and demonstrate improved convergence compared to SIRT (figure 1). The reconstruction from ten iterations of OS-EMTR still displays slight blurring around the edges, indicating it has yet to converge. Note that each iteration of OS-EMTR requires two projection operations, and so is roughly half again as computationally costly as SIRT.

## 3. PRECONDITIONED MULTIGRID RECONSTRUCTION

Having established a baseline in terms of the performance and convergence speed of existing methods, we now present our method for iterative tomographic reconstruction by preconditioned multigrid Landweber iteration.

The central idea behind multigrid iterative reconstruction is to exploit the scaling of computational complexity with the number of voxels $N$. As projection and back-projection scale strongly with $N$, they are very quick to perform on downscaled datasets. Thus in multigrid iterative reconstruction the problem is solved on a hierarchy of discrete grids, so that the quick-to-compute low-resolution solution may be used as a starting point to compute a higher-resolution solution.[13,18–20] A single iteration of the full multigrid algorithm may be defined recursively as algorithm 1. Preconditioned Landweber iteration proceeds according to

$$\mu^{(n+1)}(\mathbf{x}) = \mu^{(n)}(\mathbf{x}) + \alpha p * \mathcal{B}\left\{\ln\left[\frac{I_0(\mathbf{r},\Theta)}{I(\mathbf{r},\Theta)}\right] - [\mathcal{P}\mu^{(n)}](\mathbf{r},\Theta)\right\}, \quad (4)$$

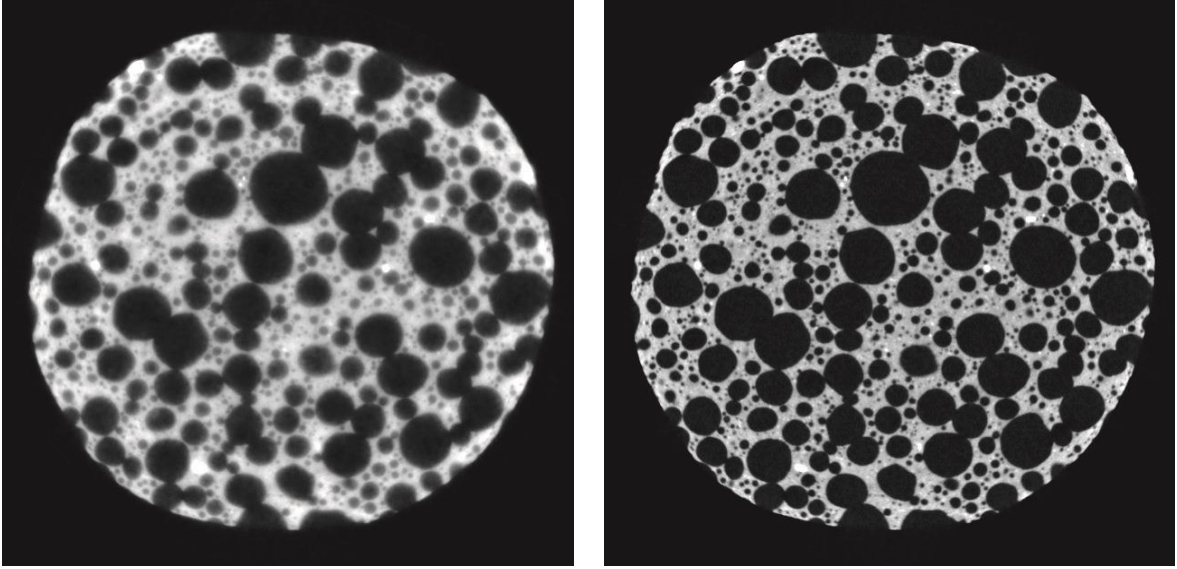

Figure 2. A $640^2$ voxel slice through the 3d OS-EMTR reconstruction of a geopolymer sample, taken orthogonal to the rotation axis, after 2 (left) and 10 (right) iterations. Note the faster convergence, compared to figure 1; this is predominantly due to the use of ordered subsets.

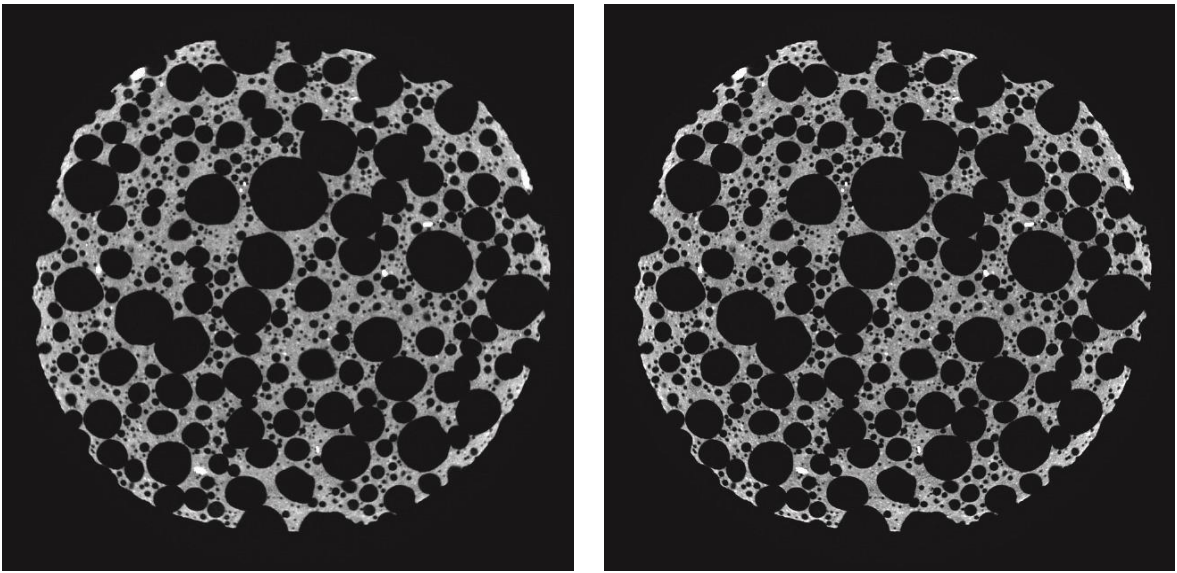

Figure 3. A $640^2$ voxel slice through the 3d preconditioned multigrid Landweber reconstruction of a geopolymer sample, taken orthogonal to the rotation axis, after 1 (left) and 2 (right) iterations. Each iteration has a computational load comparable to a single iteration of SIRT. Note that the algorithm has converged much faster than either figures 1 or 2, where the algorithms were run for 2 and 10 iterations respectively.

where $*$ denotes 3d convolution and $p$ is some preconditioning kernel chosen such that the condition number of the operator $p*\text{BP}$ is less than the condition number of the operator BP. As $p$ is a convolution kernel in the 3d volume space, the maximal reduction in condition number can be achieved when the PSF of the operator BP does not vary significantly throughout the reconstruction volume. This is the case when using a space-filling

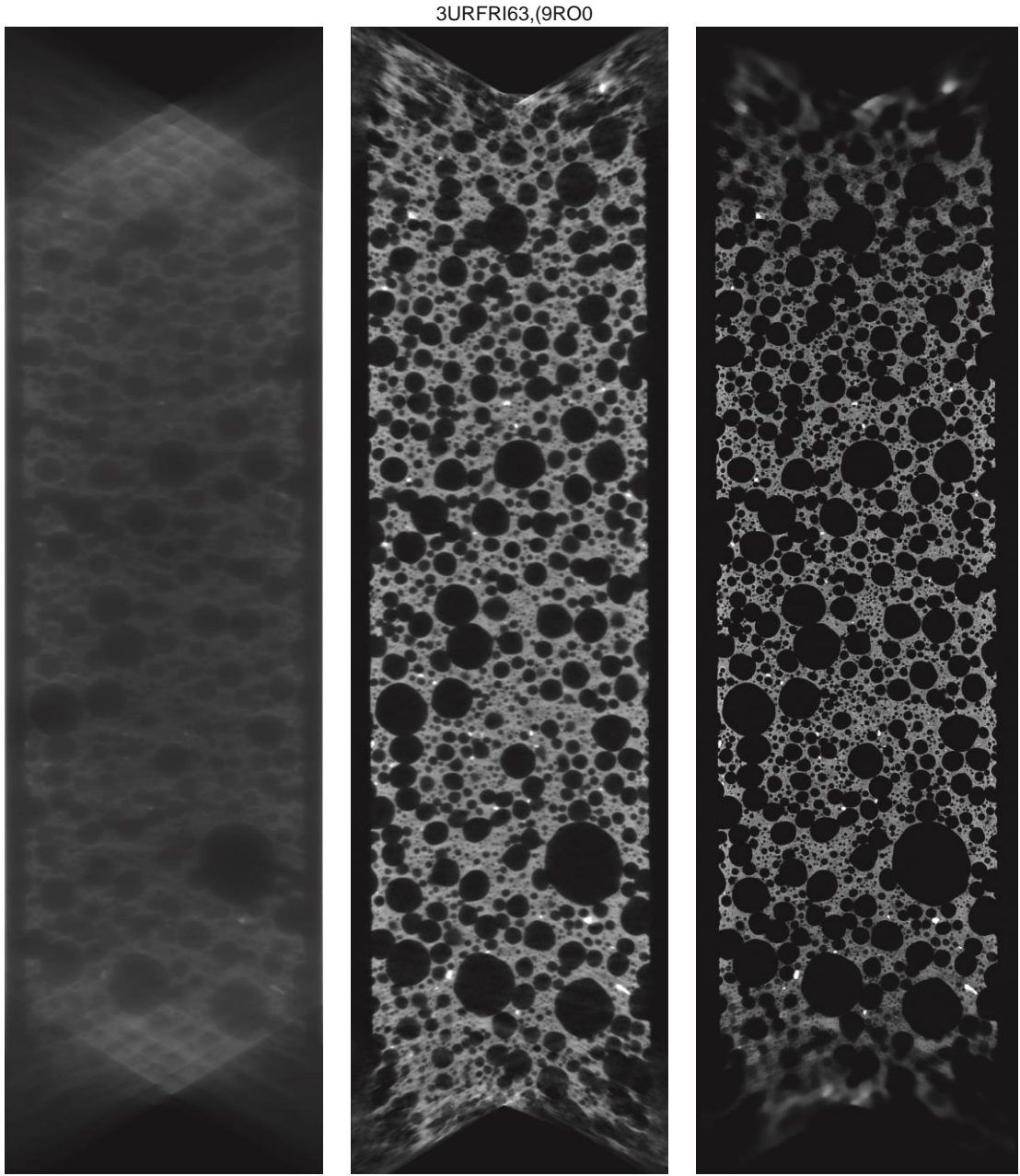

Figure 4. Vertical 640×2304 voxel slices though the geopolymer sample, after: (left) 2 iterations of SIRT; (middle) 2 iterations of OSEMTR; and (right) 2 iterations of preconditioned multigrid Landweber. The preconditioned Landweber method clearly produces better results, for the same computational cost. Note that the reconstruction degrades towards the top and bottom of the volume, as we near the end of the trajectory and the data sufficiency conditions are no longer satisfied.

trajectory; using the empirically-determined $3^3$ voxel preconditioner

$$p = \frac{1}{64} \left[ \begin{bmatrix} -1 & -1 & -1 \\ -1 & 0 & -1 \\ -1 & -1 & -1 \end{bmatrix}, \begin{bmatrix} -8 & -4 & -8 \\ -4 & 64 & -4 \\ -8 & -4 & -8 \end{bmatrix}, \begin{bmatrix} -1 & -1 & -1 \\ -1 & 0 & -1 \\ -1 & -1 & -1 \end{bmatrix} \right]$$

(5)

```
def Multigrid(μ^(n,g), I^(g), Θ): /* Recursive multigrid iterator                                    */
```
**Data**: Current estimate $\mu^{(n,g)}$ on grid $g$; radiograph intensity $I^{(g)}$ on grid $g$; scanning trajectory Θ; preconditioning filter $p$.

**Result**: Updated 3d linear attenuation coefficient $\mu^{(n+1,g)}$ on grid $g$.

/* Solve the reconstruction problem on a coarse grid $\mu^{(n,g+1)}, I^{(g+1)} \leftarrow$                    */

Restrict($\mu^{(n,g)}$, $I^{(g)}$) **if** $g <$ max:

  $\mu^{(n+1,g+1)} \leftarrow$ Multigrid($\mu^{(n,g+1)}$, $I^{(g+1)}$, Θ, $p$)

**else:**

  $\mu^{(n+1,g+1)} \leftarrow$ osemtr($\mu^{(n,g+1)}$, $I^{(g+1)}$, Θ)

/* Use the coarse-grid estimate to correct the fine grid estimate                                    */
$\mu^{(n,g)} \leftarrow \mu^{(n,g)} +$ Prolongate($\mu^{(n+1,g+1)} - \mu^{(n,g+1)}$)

/* Perform a single preconditioned Landweber iteration $\mu^{(n+1,g)} \leftarrow$                    */
Landweber($\mu^{(n,g)}$, $I^{(g)}$, Θ)

/* Apply L1 regularisation by soft thresholding                                                     */
$\mu^{(n+1,g)} \leftarrow$ SoftThreshold($\mu^{(n+1,g)}$)                                                  */

/* Apply local regularisation with a bilateral filter $\mu^{(n+1,g)} \leftarrow$
BilateralFilter($\mu^{(n+1,g)}$)

**Algorithm 1:** Recursive multigrid iterative tomographic reconstruction algorithm

results in the reconstructions shown in figure 3. These results demonstrate that the algorithm converges within one or two full-resolution iterations, each of which is approximately as computationally costly as a single iteration of SIRT.

A comparison of vertical slices (i.e. slices parallel to the source rotation axis) through the reconstructed volume after 2 iterations of SIRT, OS-EMTR, and preconditioned multigrid Landweber, is shown in figure 4. Upon visual inspection of these slices, it is clear that our preconditioned multigrid Landweber method converges much more rapidly than either SIRT or OS-EMTR. Furthermore, upon comparison of figures 2 and 3, we note that the 2-iteration preconditioned multigrid Landweber reconstruction is sharper than the 10-iteration OS-EMTR reconstruction. This is unsurprising, as the OS-EMTR algorithm continues to converge after 10 iterations.

It is important to note that the grey levels in figure 4 are not normalised between the OS-EMTR and preconditioned multigrid Landweber reconstructions. Algorithm 1 makes use of a soft-thresholding operation that progressively reduces the grey levels in the reconstruction. This is not corrected by our preconditioned Landweber iteration, as we have chosen a preconditioner with zero response to a DC offset. Thus, it is clear that a more sophisticated preconditioning kernel is required. A publication presenting more rigorously-derived and effective kernels is currently in preparation.

## 4. CONCLUSION

Figures 3 and 4 clearly demonstrate that by leveraging the approximately spatially-invariant PSF of a spacefilling trajectory, the preconditioned multigrid Landweber method converges much faster than either SIRT or OS-EMTR. It approaches a reasonable image quality within one or two iterations, with a computational cost closer to that of filtered backprojection, than to the other iterative methods we tested. Additional research has verified the scaling of this algorithm to large datasets, demonstrating a successful reconstruction of a 186 GigaVoxel volume in 16.15 hours.[12]

## ACKNOWLEDGMENTS

The authors wish to acknowledge Grzegorz Pyka at FEI Co., for collecting the experimental data used in this paper. This research was supported under the Australian Research Council's Linkage Projects funding scheme (project number LP150101040), in collaboration with FEI.